# Magnetic vortex nucleation/annihilation in artificial-ferrimagnet microdisks

Pavel N. Lapa,[1,2] Junjia Ding,[1] Charudatta Phatak,[1] John E. Pearson,[1] J. S. Jiang,[1] Axel Hoffmann[1] and Valentine Novosad[1]

[1]*Material Science Division, Argonne National Laboratory, Argonne, IL 60439, USA*

[2]*Department of Physics and Astronomy, Texas A&M University, College Station, TX 77843-4242, USA*

The topological nature of magnetic-vortex state gives rise to peculiar magnetization reversal observed in magnetic microdisks. Interestingly, magnetostatic and exchange energies which drive this reversal can be effectively controlled in artificial ferrimagnet heterostructures composed of rare-earth and transition metals. $[Py(t)/Gd(t)]_{25}$ ($t$=1 or 2 nm) superlattices demonstrate a pronounced change of the magnetization and exchange stiffness in a 10–300 K temperature range as well as very small magnetic anisotropy. Due to these properties, the magnetization of cylindrical microdisks composed of these artificial ferrimagnets can be transformed from the vortex to uniformly-magnetized states in a permanent magnetic field by changing the temperature. We explored the behavior of magnetization in 1.5-µm $[Py(t)/Gd(t)]_{25}$ ($t$=1 or 2 nm) disks at different temperatures and magnetic fields and observed that due to the energy barrier separating vortex and uniformly-magnetized states, the vortex nucleation and annihilation occur at different temperatures. This causes the temperature dependences of the Py/Gd disks magnetization to demonstrate unique hysteretic behavior in a narrow temperature range. It was discovered that for the $[Py(2\text{ nm})/Gd(2\text{ nm})]_{25}$ microdisks the vortex can be metastable at a certain temperature range.

## I. INTRODUCTION

In order to minimize magnetic stray field and magnetostatic energy associated with it, the magnetization of a macroscopic ferromagnetic object typically develops an inhomogeneous domain structure. For nano-sized objects, the cost of exchange energy due to the domain walls makes multi-domain state energetically unfavorable. Hence, for these objects, a uniformly-magnetized single-domain state is stable, even in a zero magnetic field. For cylindrical microdisks, which have sizes in between macro- and nano-size regimes, another mechanism of a magnetic-flux elimination can be realized. Namely, in a zero magnetic field, magnetization curls azimuthally around the disk geometrical center, forming a so-called magnetic vortex.[1] To avoid discontinuity of magnetization in the geometrical center of the vortex, the magnetization smoothly points perpendicular to the disk plane within a narrow area containing the vortex center. The area with a nonzero out-of-plane component of magnetization is usually called a vortex core.[2] An increasing magnetic field causes the vortex to shift away from the geometrical center of the disk in the direction perpendicular to the magnetic field. When the magnetic field exceeds a critical value, which is defined by microscopic parameters of the material (exchange stiffness and magnetization) and geometrical parameters of the disks (diameter and height), the vortex annihilates. Vice versa, a decreasing magnetic field drives the vortex nucleation and it subsequently moves towards the geometrical center of the disk.

It is argued that the drastically reduced dipole-dipole interaction between the disks and the topological nature of the vortex state can be utilized for storing binary information.[3, 4] For this, the static[5] and dynamic[6-8] behavior of magnetic vortices have been studied extensively. More fundamental aspects, like a magnetic structure of a vortex core or microscopic mechanisms of vortex nucleation/annihilation, have also attracted a lot of attention and have been investigated using a variety of tools.[9-15] Thus, fabrication of multilayered microdisks in which different layers are either coupled to each other[16, 17] or exchange biased with an antiferromagnet[18] has proved to be an efficient way to control magnetization reversal.

One of the main requirement for an observation of a vortex state in a magnetic microdisk is that the magnetic materials have very low magnetic anisotropy.[19] A typical choice of materials which satisfy this requirement includes magnetically soft Permalloy (Py=$Ni_{0.81}Fe_{0.19}$),[20] polycrystalline Ni[21] and Fe.[15] For microdisks composed of these materials, a conventional approach to switch the magnetization from the vortex to uniformly-magnetized states is to change the amplitude of the externally-applied magnetic field. Since the Curie temperatures for these materials are high, their magnetic parameters (magnetization and exchange stiffness) are almost constant below room temperature. This means that for the microdisks composed of these materials, the nucleation and annihilation fields demonstrate an insignificant change while temperature is varied within a 10–300 K range.[14]

Due to an antiferromagnetic coupling between rare-earth Gd and transition metals, the heterostructures composed of these materials are widely used for artificial ferrimagnet applications.[22-27] As we observed previously,[26, 27] if an artificial ferrimagnet is composed of thin layers of Py and Gd, the heterostructure has very low anisotropy which is comparable with that of Py. Hence, it is possible that the magnetic vortex may be a stable magnetization configuration for the microdisks composed of these artificial ferrimagnets. Importantly, the Curie temperature of Gd (293 K for bulk) is much lower than that for Py which means that the Gd/Py superlattices have significant changes of magnetization between 10 and 300 K. This suggests that the vortex nucleation and annihilation fields for the Py/Gd superlattice microdisks may also vary significantly within this temperature range. Our motivation for this work was to detect and study vortex nucleation/annihilation transitions that occur in a permanent magnetic field due to the change of the microscopic parameters of the material for different temperatures.

## II. EXPERIMENTAL DETAILS

We prepared two groups of disk arrays for the study. The disks from the first and second groups have [Py(1 nm)/Gd(1 nm)]$_{25}$ and [Py(2 nm)/Gd(2 nm)]$_{25}$ structures along the height, respectively. The nominal diameter of each disk is 1.5 μm. In addition, two different approaches were used for fabrication of each group: etching and lift-off. For fabrication of the lift-off technique, Si/SiO$_2$ wafers were spin-coated with photoresist, and an array of 1.5 μm holes were patterned by means of optical lithography. Then, after magnetron sputtering of the multilayers, the lift-off process yielded arrays of 1.5 μm disks.

For fabrication of the etched disks, the artificial ferrimagnet films were sputtered on top of the Si/SiO$_2$ wafer. After that, the arrays of holes were patterned on top of the films. This pattern was transferred into arrays of photoresist disks using an image reversal technique (baking the wafer at 100ºC in NH$_3$ environment for 30 minutes followed by flood exposure and development). The final step of this process was Ar-ion milling. The deposition rates used for the sputtering of Gd and Py are 1.4 Å/sec and 0.7 Å/sec, respectively. For seeding and a capping, 5-nm thick Ta layers were deposited before and after sputtering each multilayer. In addition to the disk arrays, two control films with the same structures were sputtered for the study. The wafers containing fabricated disk arrays were cut into 6 mm×6 mm pieces, and their magnetic behavior was studied using a superconducting quantum interference device (SQUID) magnetometer. We also fabricated the [Py(1 nm)/Gd(1 nm)]$_{25}$ disk array on top of a Si$_3$N$_4$ membrane window for a Lorentz transmission electron microscopy (TEM) study using the lift-off technique. TEM images at different temperatures were taken in Fresnel imaging mode.[28]

### III. MATERIAL PROPERTIES AND SIMULATIONS DETAILS

Magnetic and structural properties of the Py/Gd multilayers have been studied by us[26, 27] and other groups[25, 29, 30] previously. Importantly for this work, a polycrystalline Gd film has comparatively high anisotropy,[31] and only the Py/Gd multilayers with Gd-layer thickness of 2 nm or less are magnetically soft. At the same time, it was demonstrated that the influence of material intermixing becomes significant for multilayers with thin Py and Gd layers.[26, 27] The temperature dependences of magnetization for the [Py($t$)/Gd($t$)]$_{25}$ ($t$=1 or 2 nm) control films are shown in Fig. 1(a). It is seen that only the magnetization for the film with $t$=2 nm demonstrates ferrimagnet-like behavior. Its magnetization is small at room temperature. Then when the Gd layers magnetizations begin to rise the total magnetization decreases; at around 172 K, the magnetic moments of the Py and Gd layers compensate each other. At low temperatures, the total magnetization in the multilayer is Gd-aligned. In contrast, the [Py(1 nm)/Gd(1 nm)]$_{25}$ film becomes ferromagnetic only at around 275 K, which is below the Curie temperatures of bulk Gd and Py, and then its magnetization rises monotonically with decreasing temperature. Due to the strong intermixing of Py and Gd, the [Py(1 nm)/Gd(1 nm)]$_{25}$ superlattice can be considered as a homogeneous film composed of a PyGd alloy. We adopted this model for micromagnetic simulations of magnetization behavior for the [Py(1 nm)/Gd(1 nm)]$_{25}$ disks. The analysis[26, 27] showed that the magnetic and atomic structures of the [Py(2 nm)/Gd(2 nm)]$_{25}$ superlattice is more complicated. According to our estimates,[26, 27] only the 0.5-nm thick core parts of the Py and Gd layers are not subjected to intermixing, while the rest of the film is the PyGd alloy. Taking into account this magnetization profile over the [Py(2 nm)/Gd(2 nm)]$_{25}$ film thickness would lead to a very complicated micromagnetic model which would require a very small mesh, and hence, long calculation time. To avoid this, similarly to the micromagnetic model used for the [Py(1 nm)/Gd(1 nm)]$_{25}$ disks, we assumed that the [Py(2 nm)/Gd(2 nm)]$_{25}$ disks are composed of a homogeneous material. Microscopically, the effective magnetization of the Py/Gd disks, $M_{eff}$, is expected to be equal to the magnetizations of the corresponding control films at a given temperature.

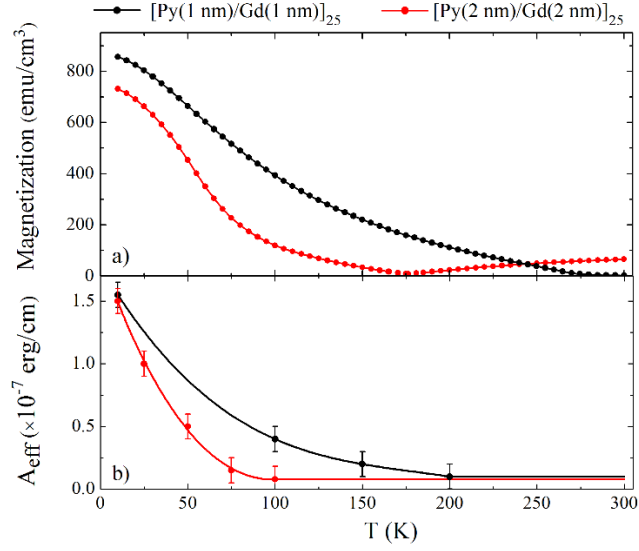

FIG. 1. a) Magnetization and b) effective exchange stiffness, $A_{eff}$, as a function of temperature for the [Py(1 nm)/Gd(1 nm)]$_{25}$ (black dots) and [Py(2 nm)/Gd(2 nm)]$_{25}$ (red dots) control films. Magnetization was measured in the 100-Oe magnetic field applied in-plane. The lines represents dependences used for the micromagnetic simulations.

Once it is assumed that the disks are composed of a homogeneous material, the effective exchange stiffness of this material at different temperatures must be estimated. For this, we utilize a technique used by us previously.[26, 27] Shortly, the films with structures Py(50 nm)/[Py($t$)/Gd($t$)]$_{25}$ ($t$=1 or 2 nm) were sputtered and their magnetization curves were measured at different temperatures. The total magnetic moment of the [Py($t$)/Gd($t$)]$_{25}$ stack is coupled antiferromagnetically with the magnetic moment of the 50-nm thick Py layer. An applied magnetic field results in an alignment of these magnetic moments along the field, and since the effective exchange stiffness of the stacks is much smaller than that of Py, the process is realized through an exchange-spring-like magnetization twist in the Py/Gd stacks. Modeling the observed non-linear dependences of the magnetization on magnetic field, and assuming that the exchange stiffness ($A_{eff}$) is constant across the Py/Gd stacks, allowed us to estimate $A_{eff}$ in the [Py($t$)/Gd($t$)]$_{25}$ ($t$=1 or 2 nm) multilayers for different temperatures [Fig. 1(b)].

Summarizing, for a micromagnetic simulation at a given temperature, $T$, it was assumed that a disk is composed of an isotropic homogeneous material with magnetization $M_{eff}(T)$ and exchange stiffness $A_{eff}(T)$. The simulations were conducted using a GPU-accelerated micromagnetic simulation program, mumax3.[32] The size of the mesh cell is 2.9 nm×2.9 nm in plane of a disk and 12.5 nm over its thickness.

## IV. EXPERIMENTAL RESULTS

For both superlattices compositions, $M_{eff}$ is at a maximum at low temperatures. Magnetization curves for the etched [Py($t$)/Gd($t$)]$_{25}$ ($t$=1 or 2 nm) disks measured at 10 K are shown in Fig. 2(a). Both curves correspond to vortex-like behavior[5] in low magnetic fields, *i.e.* the magnetization changes linearly with the field and the coercivity is negligibly small (the coercive fields are 0.5 and 2.5 Oe for $t=1$ and 2 nm, respectively). The micromagnetically-simulated curves shown in Fig. 2(b), quantitatively and qualitatively resemble the experimental ones. According to the simulations, the vortex configuration is stable in the magnetic fields of up to 340 Oe for the [Py(1 nm)/Gd(1 nm)]$_{25}$ disks, and up to 580 Oe for the [Py(2 nm)/Gd(2 nm)]$_{25}$ disks. Due to an energy barrier separating the vortex and uniformly-magnetized states, the simulated and experimental loops have hysteretic regions in higher magnetic fields [Fig. 2(a, b)]. Because the disks diameter is 1.5 μm and $A_{eff}$ is low, the vortex nucleation and annihilation are very different from those observed in submicron Py disks.[33] First, increasing magnetic field causes the magnetic vortex to deform in addition to shifting. In magnetic fields below the annihilation fields, the vortices are moon-shaped [schematics in Fig. 2(b)]. Second, upon magnetic field decrease, the magnetization does not transform from a uniform state to a vortex state directly, but instead, two vortices appear which move toward each other and merge into a single vortex in 210 and 360-Oe magnetic fields for $t=1$ and 2 nm, respectively [schematics in Fig. 2(b)]. Only the vortex annihilation looks like a sharp transition.

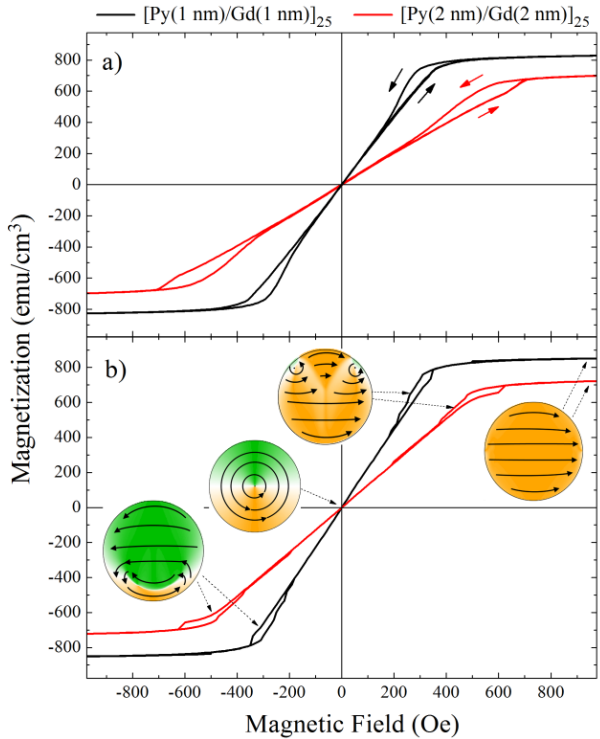

FIG. 2. a) Experimental b) micromagnetically-simulated hysteresis loops (10 K) for the etched [Py(1 nm)/Gd(1 nm)]$_{25}$ (black line) and [Py(2 nm)/Gd(2 nm)]$_{25}$ (red line) disks; b) contains the schematics of magnetization configurations realized in the disks at different magnetic fields.

To determine which magnetization configurations are realized at different temperatures, magnetization of the etched arrays was measured in a 100-Oe magnetic field while the temperature was slowly (1 K/min) swept from 300 to 10 K and back to 300 K [Fig. 3(a)]. At the initial stage of the cooling down, the curves demonstrate the behavior almost identical to that demonstrated by the control films. Then, at 112 K for $t=1$ nm and 66 K for $t=2$ nm, a phase transition occurs and the magnetization begins to decrease when the temperature is increased. At low temperatures, the magnetization is almost constant. When the temperature is ramped up, the magnetization decreases but not monotonically. For the [Py(1 nm)/Gd(1 nm)]$_{25}$ disks, the magnetization falls slowly until 143 K, and then the curve merges with the one obtained while cooling down the sample. A very similar decrease of the magnetization is demonstrated by the [Py(2 nm)/Gd(2 nm)]$_{25}$ disks in the initial stage of the warm-up, but right before the merging with the cool-down curve, the magnetization kinks up. Interestingly, the magnetic phase transitions yield the temperature dependences of the magnetization to demonstrate hysteretic behavior within the 40-160 K and 40-110 K ranges, for $t=1$ and 2 nm, respectively.

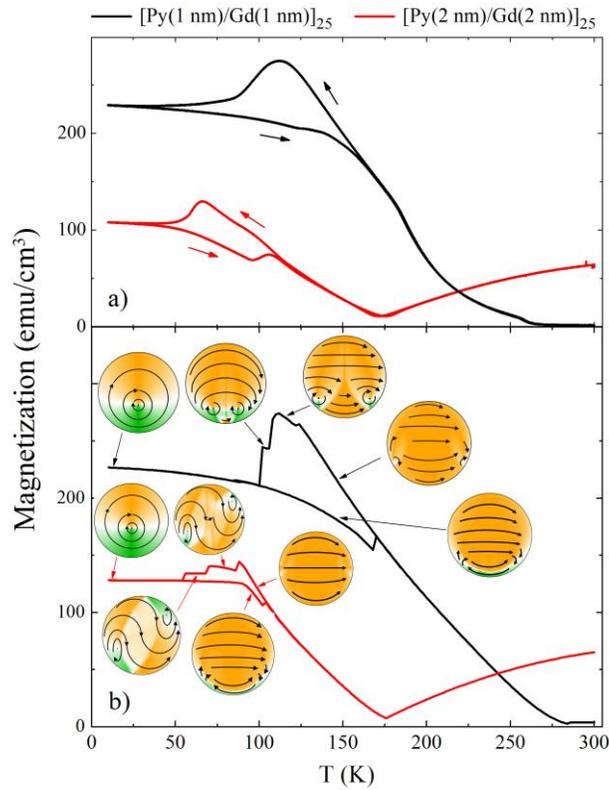

FIG. 3. a) Experimental b) micromagnetically-simulated temperature dependences of magnetization in the 100-Oe magnetic field for the etched [Py(1 nm)/Gd(1 nm)]$_{25}$ (black line) and [Py(2 nm)/Gd(2 nm)]$_{25}$ (red line) disks; b) contains the schematics of magnetization configurations realized in the disks at different temperatures.

The temperature dependences of the disks magnetization in the 100-Oe magnetic field were simulated micromagnetically [Fig. 3(b)]. Importantly, it is a zero-temperature simulations without thermal field fluctuations and any dependence on temperature is only introduced by temperature-dependent parameters, $M_{eff}(T)$ and $A_{eff}(T)$. This simplification is reasonable because any fluctuation effects, such as a thermal excitation over an energy barrier,[34] are expected to be insignificant for 1.5 μm disks. The simulated curves demonstrate hysteretic behavior similar to the one observed in the experiment [Fig. 3(a, b)]. The simulations reveal that the phase transitions observed in the experiment are due to the vortex nucleation and annihilation. At the high temperatures, the disks magnetizations are low and the disks are in the uniformly-magnetized state. When the temperature is decreased the disks magnetization increases, and at a critical temperature, it becomes energetically favorable to nucleate vortices, thus minimizing the magnetic flux. Again, similarly to the nucleation processes observed at 10 K, the disks magnetization does not switch from uniformly-magnetized to single-vortex states directly; instead, the magnetization transfers through a series of intermediate states in which the double vortices configurations are stable. The transitions between each intermediate state causes an abrupt decrease of magnetization. Since the arrays consist of disks with slightly different diameters, the abrupt transitions observed in the simulations are smeared out for the experimental curves. According to the simulations, increasing the temperature causes the vortex core shifting from the center of the disk and its deformation, eventually the disks switch to the quasi-uniformly-magnetized state [schematics in Fig. 3(b)].

It was of particular interest to investigate the stability of the vortex state at the temperatures which are within the thermal hysteresis regions, *e.g.* at 115 K for the [Py(1 nm)/Gd(1 nm)]$_{25}$ disk array and 75 K for the [Py(2 nm)/Gd(2 nm)]$_{25}$ disk array. For this, the disks were cooled down to 10 K, and the magnetic field was set to 0 Oe. This procedure allowed to ensure that the vortex is the initial magnetic state for the disks. Then, the temperature was increased to 115 K and 75 K for the [Py(1 nm)/Gd(1 nm)]$_{25}$ and for [Py(2 nm)/Gd(2 nm)]$_{25}$ disk arrays, respectively, and the arrays magnetization was measured while the magnetic field was ramped up from 0 Oe to 500 Oe and then cycled between 500 and -500 Oe [Fig. 4(a)]. The experiments were simulated micromagnetically [Fig. 4(b)]. For the [Py(1 nm)/Gd(1 nm)]$_{25}$ disk array, the magnetization shows the behavior identical to that observed for the array at 10 K, *i.e.* the magnetization rises linearly in low magnetic fields and the hysteresis is present only in the high fields. The initial part of the magnetization curve measured between 0 and 500 Oe coincides with its main part (500 Oe → -500 Oe → 500 Oe) indicating that the vortex state is stable for these disks in low magnetic field. The simulated curve is in qualitative agreement with the experimental one. In contrast, the main part of the magnetization curve for the [Py(2 nm)/Gd(2 nm)]$_{25}$ disk array has a double-waisted shape, while its initial curve demonstrates typical vortex-like behavior. Moreover, the initial magnetization curve passes outside the area under the main magnetization curve. This unusual behavior indicates that the vortex state is stable at the initial stage of the measurements, but after the saturation of the disk magnetization, the magnetization does not switch back to the vortex state while the main hysteresis loop is measured. It was also observed that, if the disk is demagnetized after measuring the main hysteresis loop, and after that, the magnetization is measured while the magnetic field is increased, the resulting curve passes within the main hysteresis loop and does not coincide with the initial magnetization curve. Thus, the (0,0) point at the M-H plane can correspond to two different magnetization states, one of which is the vortex state. This means that the vortex is a metastable state for the [Py(2 nm)/Gd(2 nm)]$_{25}$ disk at 75 K. By a metastable state, it is implied that the state cannot be accessed by isothermal changing of an external magnetic field.

Interesting, the magnetic field dependence of magnetization for the [Py(2 nm)/Gd(2 nm)]$_{25}$ disk array was not completely reproduced in the simulation [Fig. 4(b) red line]. It is possible that the fine magnetic structure within the [Py(2 nm)/Gd(2 nm)]$_{25}$ superlattice as well as its very low magnetic anisotropy must be taken into account for an adequate modeling of the magnetization reversal for the corresponding disks.

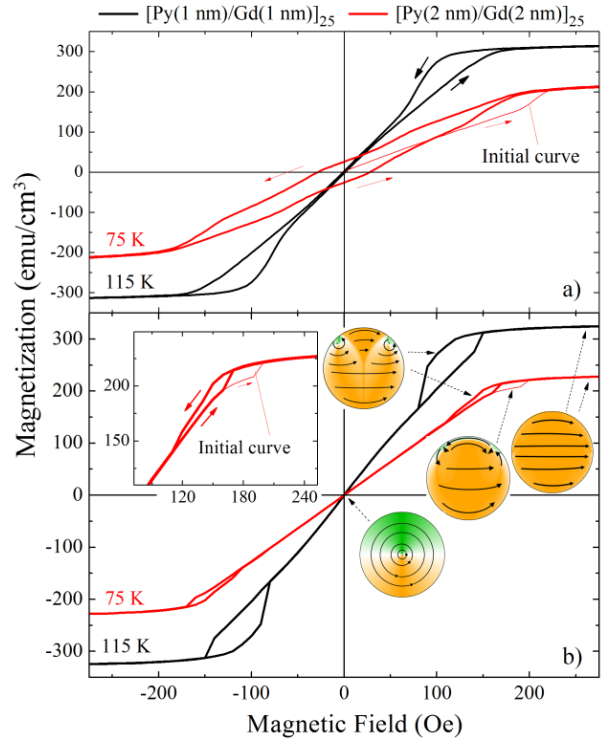

FIG. 4. a) Experimental b) micromagnetically-simulated hysteresis loops for the etched [Py(1 nm)/Gd(1 nm)]$_{25}$ disks at 115 K (black lines) and the etched [Py(2 nm)/Gd(2 nm)]$_{25}$ disks at 75 K (red lines); the initial parts of the magnetization curves (from 0 to 500 Oe) are shown with thin lines while the main parts of hysteresis curves (500 Oe → -500 Oe → 500 Oe) with thick lines; b) contains the schematics of magnetization configurations realized in the disks at different magnetic fields. The inset in the upper-left corner of (b) shows the hysteretic part of the loop for the [Py(2 nm)/Gd(2 nm)]$_{25}$ disks.

An important factor that strongly affects magnetic properties of the Py/Gd superlattice disks is the intermixing of Py and Gd. Due to a shadow effect, the intermixing is much more pronounced for the lift-off disks than that for the etched disks. Since the intermixing is almost complete for the [Py(1 nm)/Gd(1 nm)]$_{25}$ superlattice, the lift-off and etched [Py(1 nm)/Gd(1 nm)]$_{25}$ disks demonstrate almost identical temperature and field dependences. At the same time, due to the intermixing, the compensation temperature of the lift-off [Py(2 nm)/Gd(2 nm)]$_{25}$ disks is about 50 K higher than that for the etched disks and the corresponding control films. Although the same physical phenomena are observed in the lift-off [Py(2 nm)/Gd(2 nm)]$_{25}$ disks as that observed in the etched ones (vortex nucleation and annihilation), some quantitative parameters, like the nucleation/annihilation fields and temperatures, are different for the [Py(2 nm)/Gd(2 nm)]$_{25}$ disks prepared using different fabrication techniques.

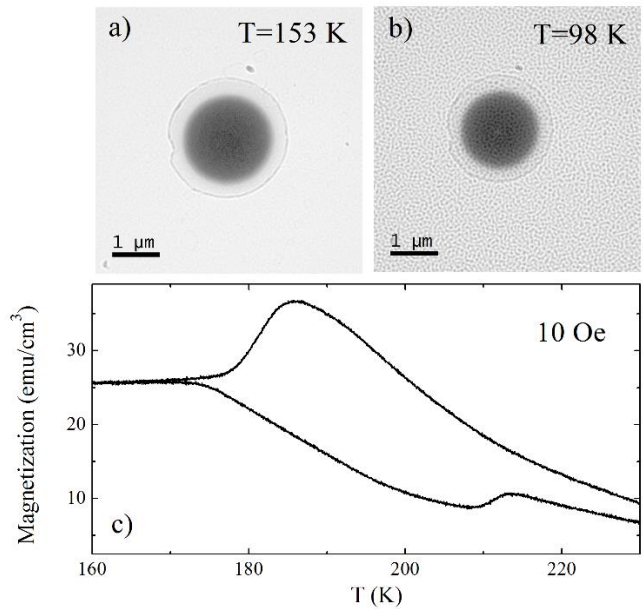

FIG. 5. Under-focused Lorentz transition electron microscopy images of a lift-off [Py(1 nm)/Gd(1 nm)]$_{25}$ disk on a Si$_3$N$_4$ membrane window acquired at 153 K (a) and 98 K (b). A white dot at the center of the disk at 98 K indicates the presence of magnetic vortex; c) magnetization of the etched [Py(1 nm)/Gd(1 nm)]$_{25}$ disk array as a function of temperature measured in the 10-Oe magnetic field.

To get a direct confirmation that vortices nucleate in the Py/Gd superlattice disks at low temperatures, we acquired a series of under-focused Lorentz TEM images of a lift-off [Py(1 nm)/Gd(1 nm)]$_{25}$ disk on a Si$_3$N$_4$ membrane at different temperatures. Fig. 5(a) and (b) shows the Lorentz TEM obtained at 152 and 98 K, respectively. A white dot at the center of the disk in Fig. 5(b), which appears when the temperature drops below 129 K, indicates the presence of a magnetic vortex. Fig. 5(c) shows a temperature dependence of the magnetization measured in very low magnetic field (10 Oe), according to which the magnetization of the [Py(1 nm)/Gd(1 nm)]$_{25}$ disks switches to the vortex state below 180 K in the low magnetic field. Importantly, during the TEM imaging the temperature was swept continuously. The real temperature of the disk on the Si$_3$N$_4$ membrane can be very different from the readings of the microscope's stage thermocouple. This yields slightly different vortex nucleation temperatures detected using Lorentz TEM and using magnetometry.

## V. CONCLUSION

We observed that the 1.5-μm disks composed of the [Py(t)/Gd(t)]$_{25}$ (t=1 or 2 nm) artificial ferrimagnets experience phase transitions. For these disks, the resulting temperature dependences of magnetization measured in the 100-Oe magnetic field are hysteretic in narrow temperature regions. Micromagnetic simulations revealed that these phase transitions are due to nucleation and annihilation of magnetic vortices which happen because the magnetic properties of the artificial ferrimagnet change significantly within the 10–300 K temperature range. It was shown that at 75 K, the vortex in the [Py(2 nm)/Gd(2 nm)]$_{25}$ disk is in a metastable state. It is impossible to nucleate the vortices in the [Py(2 nm)/Gd(2 nm)]$_{25}$ disks by changing the external magnetic field isothermally, at the same time, if the disk with the magnetic vortex is brought to 75 K, the vortex does not annihilate until the disk magnetization is saturated.